\documentclass[journal]{IEEEtran}

\ifCLASSINFOpdf
\else
   \usepackage[dvips]{graphicx}
\fi
\usepackage{url}

\usepackage{color,array,amsthm}
\usepackage{algpseudocode}
\usepackage{graphicx}
\usepackage{subfigure}
\usepackage{bm}
\usepackage{textcomp}
\usepackage{amsmath}
\usepackage{lipsum}
\usepackage{cuted}\stripsep -3pt plus 3pt minus 2pt
\usepackage{amsthm}
\usepackage{amssymb}
\usepackage{amsfonts}
\usepackage{latexsym}
\usepackage{amsfonts}
\usepackage{algorithm}

\usepackage{booktabs}
\usepackage{threeparttable}
\usepackage{multirow}
\usepackage{algorithmicx}  
\usepackage{algpseudocode}

\graphicspath{ {figures/} }

\begin{document}

\title{8D Parameters Estimation for Bistatic EMVS-MIMO Radar via the nested PARAFAC}

\author{Qianpeng Xie, He Wang, Yihang Du, Xiaoyi Pan  and Feng Zhao
\thanks{ This work was supported in part by the National Natural Science Foundation of China under Grant 61890545, 61890542 and 61890540. }
\thanks{Qianpeng Xie, Xiaoyi Pan, Feng Zhao are with the State Key Laboratory of Complex Electromagnetic Environment Effects on Electronics and Information System, National University of Defense Technology, Changsha 410073, China(email:13721038905@163.com, mrpanxy@nudt.edu.cn, zhfbee@tom.com).He Wang is with the School of Electronic Science, National University of Defense Technology, Changsha 410073, China.(email:hw8g16@soton.ac.uk).Yihang Du is with the Sixty-Third Research Institute, National University of Defense Technology, Nanjing 210007, China(email:dyhcsl1991@163.com).{{ ({Corresponding author: Qianpeng Xie} )}. }}}

\markboth{xxxxx, Vol. xx, No. xx, August xxxx}
{Shell \MakeLowercase{\textit{et al.}}: Bare Demo of IEEEtran.cls for IEEE Journals}
\maketitle

\begin{abstract}
In this letter, {a novel}  nested PARAFAC algorithm was proposed to improve the 8D parameters estimation performance for the bistatic EMVS-MIMO radar. Firstly, the outer part PARAFAC algorithm was carried out to estimate the receive spatial response matrix and its first way factor matrix. For the estimated first way factor matrix, a theory is given to rearrange its data into an new matrix, which is the mode-1 unfolding matrix of a three-way tensor. Then, the inner part PARAFAC algorithm was used to estimate the transmit steering vector matrix, the transmit spatial response matrix and the  receive steering vector matrix. Thus, the transmit 4D parameters and receive 4D parameters can be accurately located via the {abovementioned} process. Compared with the original PARAFAC algorithm, the proposed nested PARAFAC algorithm  can avoid additional reconstruction process when estimating  the transmit/receive spatial response matrix. Moreover, the proposed algorithm  can {offer} a highly-accurate 8D parameters estimaiton than that of the original PARAFAC algorithm. {Simulated} results verify the effectiveness of the proposed algorithm.
\end{abstract}


\IEEEpeerreviewmaketitle

\section{Introduction}
\IEEEPARstart{R}{ecently}, the research about the electromagnetic vector sensors (EMVS)  has attracted extensive attention  due to its excellent measurement capabilities of angle parameters and polarization parameters  \cite{WongK}-\cite{LanX}. In order to estimate the polarization parameters in bistatic EMVS-MIMO radar,  the ESPRIT algorithm {was} firstly proposed in \cite{Chintagunta}. The significance of the ESPRIT algorithm is to derive a method to reconstruct the transmit/receive spatial response matrix, but its computational complexity is large. In order to reduce the computational complexity, the PM algorithm  was proposed {in \cite{LiuT}}, which can estimate the signal subspace matrix by {avoiding}  the singular value decomposition process of the covariance matrix. To make full use of the space-time multidimensional structure of the array received data, the HOSVD algorithm based on the fourth-order tensor was {raised} in \cite{MaoC}. However, the algorithms {reported} in \cite{Chintagunta}-\cite{MaoC} need the pair matching process for the 2D-DOD and 2D-DOA estimation.
Therefore, a modified PM algorithm in \cite{WenF} and PARAFAC algorithm in \cite{WenFShi}  were proposed  to realize the 2D-DOD and 2D-DOA automatically paired. In \cite{{XianpengWang}}, the bistatic coprime EMVS-MIMO radar  {was} designed to improve the 2D-DOD and 2D-DOA estimaiton performance.
\par 
Through analysis, it can be found that the above mentioned algorithms can be used to estimate the 8D parameters in bistatic EMVS-MIMO radar. {Amongst}, the HOSVD algorithm can offer the best  estimation performance{, but its shortcoming} is the high computational load. The modified PM algorithm and the PARAFAC algorithm can realize the 2D-DOD and 2D-DOA automatically paired. But, the estimation performance of them are relatively poor. According to \cite{Almeida}-\cite{Paulo}, the nested PARAFAC algorithm can improve the estimation performance of the angle parameter, {but, it} cannot be directly used in the bistatic EMVS-MIMO radar due to the special structure of the receive data.
Motivated by the strategy in \cite{Almeida}-\cite{Paulo},  we {proposed} a new nested PARAFAC algorithm. Firstly, the outer part PARAFAC {was conducted } to estimate the receive spatial response and the first way factor matrix. Then, the inner part three way  tensor can be constructed by rearranging the estimated first way factor matrix, which can be used to estimate transmit {steering} vector matrix, transmit spatial response matrix and the  receive {steering} vector matrix. Compared with the HOSVD algorithm, the {raised} new nested PARAFAC algorithm has lower computational complexity. {And this method presents a higher estimation accuracy than the original PARAFAC algorithm as well as the modified PM algorithm.}  
\section{DATA MODEL}
\begin{figure}[htbp]
	\centerline{\includegraphics[width=18.5pc]{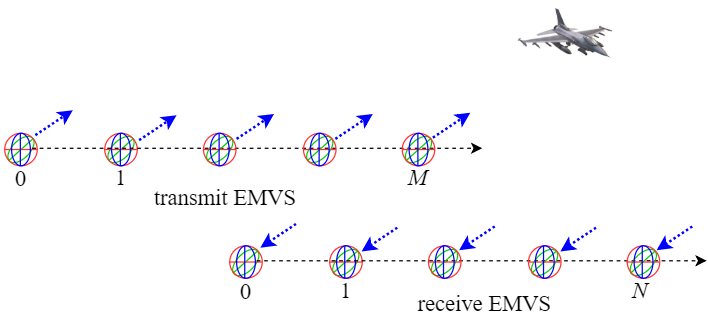}}
	\caption{Bistatic EMVS-MIMO radar system }
	\label{array1}
\end{figure}

As shown in Fig. \ref{array1}, for a bistatic EMVS-MIMO radar system {consists of} $M$ transmit EMVS and $N$  receive EMVS, the transmit steering vector and receive steering vector can be expressed as
\begin{equation}
\begin{aligned}
\boldsymbol{c}_{t_{k}} = \boldsymbol{a}_{t_{k}}\otimes \boldsymbol{q}_{t_{k}}(\theta_{t_{k}},\phi_{t_{k}},\gamma_{t_{k}},\eta_{t_{k}})
\end{aligned}    
\end{equation}
\begin{equation}
\begin{aligned}
\boldsymbol{c}_{r_{k}} = \boldsymbol{a}_{r_{k}}\otimes \boldsymbol{q}_{r_{k}}(\theta_{r_{k}},\phi_{r_{k}},\gamma_{r_{k}},\eta_{r_{k}})
\end{aligned}   \label{10}
\end{equation}
where {$k=1,2,\dots,K$ } denote the number of targets,  $\theta_{t_{k}},\theta_{r_{k}}\in [0,\pi )$, $\phi_{t_{k}},\phi_{r_{k}}\in [0,2\pi )$, $\gamma_{t_{k}},\gamma_{r_{k}}\in [0,\pi/2 )$ and $\eta_{t_{k}},\eta_{r_{k}}\in [-\pi,\pi )$  denote the transmit/receive elevation angle, transmit/receive azimuth angle, transmit/receive polarization angle and transmit/receive polarization phase difference, respectively.  $\boldsymbol{q}_{t_{k}}(\theta_{t_{k}},\phi_{t_{k}},\gamma_{t_{k}},\eta_{t_{k}})$ and $\boldsymbol{q}_{r_{k}}(\theta_{r_{k}},\phi_{r_{k}},\gamma_{r_{k}},\eta_{r_{k}}) $ represent the spatial response of the transmit EMVS and the receive EMVS of the $k$-th target.  ${\boldsymbol{a}_{t_{k}}=}{[
	e^{-j\pi\sin(\theta_{t_{k}})},e^{-j\pi2 \sin(\theta_{t_{k}})},\cdots,e^{-j\pi M \sin(\theta_{t_{k}})}
	]^{T}}$,  ${\boldsymbol{a}_{r_{k}}}{=}$ ${[
	e^{-j\pi \sin(\theta_{r_{k}})}, e^{-j\pi 2 \sin(\theta_{r_{k}})}, \cdots, e^{-j\pi N \sin(\theta_{r_{k}})}
	]^{T}}$. Accodring to \cite{WongK}, the spatial response $\boldsymbol{q}$ of an EMVS can be expressed as
\begin{equation}
\begin{aligned}
&\boldsymbol{q}(\theta,\phi,\gamma,\eta)        
= \\
&\underbrace{\begin{bmatrix}
	\cos(\phi)\cos(\theta) & -\sin(\phi)\\
	\sin(\phi)\cos(\theta)  &\cos(\phi)  \\
	-\sin(\theta)  & 0 \\
	- \sin(\phi)   & -\cos(\phi)\cos(\theta)\\
	\cos(\phi) &   -\sin(\phi)\cos(\theta) \\
	0 &    \sin(\theta)\\
	\end{bmatrix} }_{\boldsymbol{F}(\theta,\phi)}   
\underbrace{\begin{bmatrix}
	\sin(\gamma)e^{j\eta} \\
	\cos(\gamma) \\
	\end{bmatrix}}_{\boldsymbol{g}(\gamma,\eta} 
\end{aligned}    \label{11}
\end{equation}
where ${\boldsymbol{F}(\theta,\phi)}\in \mathbb{C}^{6\times 2}$ denotes the spatial angular location matrix , ${\boldsymbol{g}(\gamma,\eta)} \in \mathbb{C}^{2\times 1}$  denotes the polarization states vector.
\par The data model  at the receive front  can be expressed as
\begin{equation}
\begin{aligned}
\boldsymbol{y}(t)&=\left [ (\boldsymbol{A}_{t} \odot \boldsymbol{Q}_{t} )\odot (\boldsymbol{A}_{r} \odot \boldsymbol{Q}_{r} )  \right ] \boldsymbol{s}(t) + \boldsymbol{n}(t)\\
\end{aligned}   
\end{equation}
where $\boldsymbol{A}_{t}=
\left [ \boldsymbol{a}_{t_{1}},\boldsymbol{a}_{t_{2}},\cdots,\boldsymbol{a}_{t_{K}} \right ] 
$, $\boldsymbol{A}_{r}=
[ \boldsymbol{a}_{r_{1}},\boldsymbol{a}_{r_{2}},\cdots,$ $\boldsymbol{a}_{r_{K}}] 
$, $\boldsymbol{Q}_{t}=
\left [ \boldsymbol{q}_{t_{1}} ,\boldsymbol{q}_{t_{2}},\cdots,\boldsymbol{q}_{t_{K}} \right ] 
$ and $\boldsymbol{Q}_{r}=
 [\boldsymbol{q}_{r_{1}} ,\boldsymbol{q}_{r_{2}},$ $\cdots,\boldsymbol{q}_{r_{K}} ] 
$ denote the transmit steering  matrix, the receive steering  matrix, the transmit  spatial response matrix  and the receive spatial response matrix, respectively. For the total $L$ snapshots, the sample data can be further expressed  {written} as
\begin{equation}
\begin{aligned}
\boldsymbol{Y}
&=\left [ (\boldsymbol{A}_{t} \odot \boldsymbol{Q}_{t} )\odot (\boldsymbol{A}_{r} \odot \boldsymbol{Q}_{r} )  \right ] \boldsymbol{S} + \boldsymbol{N}
\end{aligned}   
\end{equation}
\section{The proposed  strategy}
\subsection{The nested PARAFAC algorithm}
\par In order to reserve the mutiple space-time structure in $\boldsymbol{Y}$, a nested-tensor signal model are designed as follows 
\begin{equation}
	\begin{aligned}
		\boldsymbol{Y}
		&=\left [ \left ( \boldsymbol{A}_{t} \odot \boldsymbol{Q}_{t} \right )  \odot \left ( \boldsymbol{A}_{r} \odot \boldsymbol{Q}_{r} \right ) \right ]    \boldsymbol{S} +  \boldsymbol{N}\\
		&= \underbrace{\left [ \overbrace{\left ( \boldsymbol{A}_{t} \odot \boldsymbol{Q}_{t}\odot \boldsymbol{A}_{r} \right ) }^{inner \quad part}  \odot \boldsymbol{Q}_{r} \right ] \boldsymbol{S}+  \boldsymbol{N}}_{outer \quad part}   \\
		&= \left [ \boldsymbol{A}_{tqr} \odot \boldsymbol{Q}_{r} \right ]  \boldsymbol{S}+  \boldsymbol{N}
	\end{aligned}  \label{6} 
\end{equation}
where $\boldsymbol{A}_{tqr}=\left ( \boldsymbol{\tilde{A}}_{t} \odot \boldsymbol{Q}_{t}\odot \boldsymbol{\tilde{A}}_{r} \right )\in \mathbb{C}^{6MN\times K}$. Firstly, the outer part PARAFAC algorithm is  used to estimate the factor matrices $\boldsymbol{A}_{tqr}$ and $\boldsymbol{Q}_{r}$. According to (\ref{6}), a 3-D tensor $\mathcal{Y}$ can be constructed as
\begin{equation}
\begin{aligned}
\mathcal{Y}
&= \sum_{k=1}^{K} \boldsymbol{a}_{tqr}\circ \boldsymbol{q}_{r} \circ \boldsymbol{s}^{T} + \mathcal{N}
\end{aligned}   
\end{equation}
\par  The different silices of  the  third-order tensor $\mathcal{Y}$ can be further expressed as 
\begin{equation}
\begin{aligned}
\mathcal{Y}^{T}_{[{i,:,:}]} &=  \boldsymbol{Q}_{r}\mathcal{D}_{i}( {\boldsymbol{S}}^{T}) \boldsymbol{A}^{T}_{tqr} ,\quad i =1,2,\cdots,L\\
\mathcal{Y}^{T}_{[{:,j,:}]} &=\boldsymbol{S}^{T}\mathcal{D}_{j}({{\boldsymbol{{A}_{tqr}}}} )  \boldsymbol{Q}_{r}^{T}  ,\quad j=1,2,\cdots,6{MN}\\
\mathcal{Y}^{T}_{[{:,:,k}]} &= \boldsymbol{A}_{tqr} \mathcal{D}_{k}(\boldsymbol{Q}_{r})\boldsymbol{S}  ,\quad k=1,2,\cdots,6\\
\end{aligned} \label{8}  
\end{equation}
where $\mathcal{D}_{i}$, $\mathcal{D}_{j}$, $\mathcal{D}_{k}$ represent the diagonal matrix operation, and the elements on the diagonal matrices are the $k$-rows of the factor matrices $\boldsymbol{Q}_{r}$, $\boldsymbol{A}_{tqr}$, and ${\boldsymbol{S}}$, respectively.
\par In order to realize the estimation of the factor matrices $\boldsymbol{Q}_{r}$, $\boldsymbol{A}_{tqr}$, and ${\boldsymbol{S}}$, the trilinear alternating least squares algorithm is used 
\begin{equation}
\begin{aligned}
\min_{{\boldsymbol{A}^{T}_{tqr}}}&=\Vert \lbrack \mathcal{Y} \rbrack^{T}_{(1)} -    \left ( \boldsymbol{Q}_{r}\odot {\boldsymbol{S}}^{T} \right )  \boldsymbol{A}^{T}_{tqr}  \Vert^{2}_{F}\\
\min_{{\boldsymbol{Q}^{T}_{r}}}&=\Vert \lbrack  \mathcal{Y}\rbrack^{T}_{(2)} -   \left ( \boldsymbol{S}^{T}\odot  {{\boldsymbol{{A}_{tqr}}}}  \right )  \boldsymbol{Q}_{r}^{T}      \Vert^{2}_{F}\\
\min_{{\boldsymbol{S}}}&=\Vert \lbrack  \mathcal{Y}\rbrack^{T}_{(3)} -  \left ( \boldsymbol{A}_{tqr} \odot  \boldsymbol{Q}_{r} \right )  \boldsymbol{S}               \Vert^{2}_{F}\\
\end{aligned}   \label{9} 
\end{equation}
\par Then, let ${{\boldsymbol{\breve{S}}}}$, ${\boldsymbol{\breve{Q}}_{r}}$ and ${\boldsymbol{\breve{A}}_{tqr}}$ denote the estimated factor matrices, respectively
\begin{equation}
\begin{aligned}
&{\boldsymbol{\breve{A}}^{T}_{tqr}}= ({\boldsymbol{\breve{Q}}_{r}} \odot {\boldsymbol{\breve{S}}^{T}})^{\dagger}\lbrack  \mathcal{Y}\rbrack^{T}_{(1)} \\
&{{\boldsymbol{\breve{Q}}^{T}_{r}}}= ({\boldsymbol{\breve{S}}}^{T}\odot {\boldsymbol{\breve{A}}_{tqr}})^{\dagger}\lbrack  \mathcal{Y}\rbrack^{T}_{(2)} \\
&{{\boldsymbol{\breve{S}}}} = (\boldsymbol{\breve{A}}_{tqr}\odot {\boldsymbol{\breve{Q}}_{r}})^{\dagger}\lbrack  \mathcal{Y} \rbrack^{T}_{(3)} \\
\end{aligned}   \label{10}
\end{equation}
\par  It can be found that the factor matrices ${\boldsymbol{\breve{Q}}_{r}}$ and ${\boldsymbol{\breve{A}}_{tqr}}$ can be obtained. And, the receive elevation angle, receive azimuth angle, receive polarization angle and receive polarization phase difference can be estimated by performing the vector-cross-product operation on ${\boldsymbol{\breve{Q}}_{r}}$.
\par Then, in order to estimate the transmit 4D parameters, the inter part in (\ref{6}) {needs to be considered}. {It is invalid to} direct to perform inner-part PARAFAC on estimated ${\boldsymbol{\breve{A}}_{tqr}}$.  {Under this circumstance, a new three way tensor should be constructed as follows}
\begin{equation}
\begin{aligned}
\mathcal{Y}_{1}
&= \sum_{k=1}^{K} \boldsymbol{A}_{r}\circ\boldsymbol{q}_{t}\circ \boldsymbol{A}_{t}   
\end{aligned}   
\end{equation}
\par {However}, the factor matrix ${\boldsymbol{\breve{A}}_{tqr}}\ne \mathcal{Y}_{1{\left ( i \right ) } },i=1,2,3$. Thus,  {the conversion of} the factor matrix ${\boldsymbol{\breve{A}}_{tqr}}$ into the mode-$i,i=1,2,3$ unfolding of the tensor $\mathcal{Y}_{1}$ { is necessary}. By analysis,  the relationship $\boldsymbol{A}_{tqr}$ and  $\mathcal{Y}_{1{\left ( 1 \right ) } } ${is considered}. The detailed form of  the mode-$1$ unfolding of the tensor $\mathcal{Y}_{1}$ is
\begin{equation}
\begin{aligned}
\widehat{\boldsymbol{A}_{tqr}} =\left [ \mathcal{Y}_{1 } \right ]^{T}_{{\left ( 1 \right ) }} =\left [  {\boldsymbol{{Q}}_{t}} \odot {\boldsymbol{{A}}_{t}}  \right ]{\boldsymbol{{A}}^{T}_{r}} \in \mathbb{C}^{6M\times N}
\end{aligned}   
\end{equation}
\par Here, we give an theory to show how {to} rearrange the factor matrix $\boldsymbol{A}_{tqr}$ into the matrix $\widehat{\boldsymbol{A}_{tqr}}$ 
\par 
${\mathit{Theorem\quad 1} }$ The matrix $\widehat{\boldsymbol{A}_{tqr}}\in \mathbb{C}^{6M\times N}$ is equal to the inverse matrix vectorization by row of the vector $\left [ \sum_{k=1}^{K}\boldsymbol{A}_{tqr}\left ( :,k \right )  \right ]\in \mathbb{C}^{6MN\times 1}  $, namely 
\begin{equation}
\begin{aligned}
\widehat{\boldsymbol{A}_{tqr}} = ivec\left [ \sum_{k=1}^{K}\boldsymbol{A}_{tqr}\left ( :,k \right )  \right ]  
\end{aligned}   
\end{equation}\\
where $ivec$ denotes the inverse matrix vectorization by row.
\par ${\mathit{Proof} }:$ The conclusion is a direct result of the following equalities
\begin{equation}
\begin{aligned}
\left [ \sum_{k=1}^{K}\boldsymbol{A}_{tqr}\left ( :,k \right )  \right ] 
&\Leftrightarrow \sum_{k=1}^{K}\left ( \boldsymbol{A}_{t}\left ( :,k \right ) \odot \boldsymbol{Q}_{t}\left ( :,k \right )\odot \boldsymbol{A}_{r} \left ( :,k \right )\right )\\
&\Leftrightarrow \sum_{k=1}^{K} \left ( \boldsymbol{a}_{tk} \otimes \boldsymbol{q}_{tk}\otimes \boldsymbol{a}_{rk} \right )\\
& \overset{1}{\Leftrightarrow}  \sum_{k=1}^{K}vec\left (\boldsymbol{a}_{rk} \left ( \boldsymbol{a}_{tk} \otimes \boldsymbol{q}_{tk} \right )^{T}  \right )\\
&\overset{2}{\Leftrightarrow} vec\left (\boldsymbol{A}_{r} \left ( \boldsymbol{A}_{t} \odot \boldsymbol{Q}_{t} \right )^{T}
\right )\\
&\Leftrightarrow vec\left (  \left [ \left (  {\boldsymbol{{Q}}_{t}} \odot {\boldsymbol{{A}}_{t}}  \right ){\boldsymbol{{A}}^{T}_{r}} \right ]^{T} 
\right )\\
&\Leftrightarrow vec\left ( \left ( \widehat{\boldsymbol{A}_{tqr}}  \right )^{T} \right )  
\end{aligned}   
\end{equation} 
where the $\overset{1}{\Leftrightarrow}$ is satisfied by using the relationship between kronecker product and vectorization of two vector  $\boldsymbol{a}\otimes  \boldsymbol{b}= \boldsymbol{a}\odot   \boldsymbol{b}=vec\left (\boldsymbol{b} \boldsymbol{a}^{T}  \right ) $. Let $\boldsymbol{{B}}={\boldsymbol{{Q}}_{t}} \odot {\boldsymbol{{A}}_{t}}=\left [\boldsymbol{{b}}_{1},\boldsymbol{{b}}_{2},\cdots,\boldsymbol{{b}}_{K}  \right ] $, 
the $\overset{2}{\Leftrightarrow}$ is equivalent by  the follwing analysis 
\begin{equation}
\begin{aligned}
vec\left ( \boldsymbol{{A}}_{r} \boldsymbol{{B}}^{T} \right ) &\Leftrightarrow vec \left ( \left [\boldsymbol{{a}}_{r1},\boldsymbol{{a}}_{r2},\cdots,\boldsymbol{{a}}_{rK}  \right ]\begin{bmatrix}
\boldsymbol{{b}}_{1} \\
\boldsymbol{{b}}_{2}\\
\cdots\\
\boldsymbol{{b}}_{K}
\end{bmatrix} \right ) \\
&\Leftrightarrow vec\left ( \boldsymbol{{a}}_{r1}\boldsymbol{{b}}_{1}+\boldsymbol{{a}}_{r2}\boldsymbol{{b}}_{2}+\dots +\boldsymbol{{a}}_{rK}\boldsymbol{{b}}_{K} \right ) \\
&\Leftrightarrow \sum_{k=1}^{K}vec\left (\boldsymbol{a}_{rk}\boldsymbol{{b}}^{T}_{k}   \right )\\
&\Leftrightarrow \sum_{k=1}^{K}vec\left (\boldsymbol{a}_{rk} \left ( \boldsymbol{a}_{tk} \otimes \boldsymbol{q}_{tk} \right )^{T}  \right )
\end{aligned}   
\end{equation}\\   
\par {Consequently}, the matrix $\widehat{\boldsymbol{A}_{tqr}}$ can be obtained by using the above theory. {Subsequently}, the inter part three-way tensor $\mathcal{Y}_{1}$ can be constructed as
\begin{equation}
\begin{aligned}
\mathcal{Y}_{1}
& =  reshape\left ( \widehat{\boldsymbol{A}_{tqr}},\left [ N,6,M \right ]  \right )=
\sum_{k=1}^{K}  \boldsymbol{A}_{r} \circ \boldsymbol{q}_{t}\circ \boldsymbol{A}_{t} 
\end{aligned}   
\end{equation}
\par Therefore, the PARAFAC algorithm can be used again to estimate the factor matrices  ${\boldsymbol{\breve{Q}}_{t}}$, ${\boldsymbol{\breve{A}}_{t}}$ and ${\boldsymbol{\breve{A}}_{r}}$. The {derivation} process is similar to  (\ref{8})-(\ref{10}). Thus, {through the utilization of} the outer part PARAFAC and inner part PARAFAC, the factor matrices ${\boldsymbol{\breve{Q}}_{r}}$, ${\boldsymbol{\breve{Q}}_{t}}$, ${\boldsymbol{\breve{A}}_{t}}$ and ${\boldsymbol{\breve{A}}_{r}}$ can be obtained.
\subsection{Estimaiton of the angle and polarization parameters}
\par 
For the estimated ${\boldsymbol{\breve{Q}}_{r}}$, ${\boldsymbol{\breve{Q}}_{t}}$, ${\boldsymbol{\breve{A}}_{t}}$ and ${\boldsymbol{\breve{A}}_{r}}$, the corresponding   transmit 4-D parameters $(\boldsymbol{\theta}_{t},\boldsymbol{\phi}_{t},\boldsymbol{\gamma}_{t},\boldsymbol{\eta}_{t})$ and the receive 4-D parameters $(\boldsymbol{\theta}_{r},\boldsymbol{\phi}_{r},\boldsymbol{\gamma}_{r},\boldsymbol{\eta}_{r})$ can be {attained}. As the derivation is similar, only the derivation for  $(\boldsymbol{\theta}_{r},\boldsymbol{\phi}_{r},\boldsymbol{\gamma}_{r},\boldsymbol{\eta}_{r})$ { is provided here.} For estimated ${\boldsymbol{\breve{A}}_{r}}$ { via}  the inner part PARAFAC algorithm, the following selection matrices {can be constructed}
\begin{equation}
	\begin{aligned}
		\begin{cases}
			\boldsymbol{J}_{r1} = \begin{bmatrix}
				\boldsymbol{I}_{N-1}  & \boldsymbol{0}_{\left ( {N-1} \right )\times 1 }
			\end{bmatrix} \\
			\boldsymbol{J}_{r2} = \begin{bmatrix}
				\boldsymbol{0}_{\left ( {N-1} \right )\times 1 } & \boldsymbol{I}_{N-1}  
			\end{bmatrix}
		\end{cases}\\ 
	\end{aligned}   
\end{equation}
\par Then,  the following rotation-invariant relationship for $\boldsymbol{\widehat{A}}_{r}$ can be {obtained as}
\begin{equation}
	\begin{aligned}
		\boldsymbol{J}_{r1}\boldsymbol{\widehat{A}}_{r}\boldsymbol{\Phi}(\theta_{r})=\boldsymbol{J}_{r2}\boldsymbol{\widehat{A}}_{r}
	\end{aligned} \label{18}  
\end{equation}
where
\begin{equation}
	\begin{aligned}
		\boldsymbol{\Phi}(\theta_{r})=\begin{bmatrix}
			e^{j\pi\sin(\theta_{r_{1}})}&&\\
			&\ddots& \\
			&&e^{j\pi\sin(\theta_{r_{K}})}\end{bmatrix}
	\end{aligned}   
\end{equation}\\
 \par {Resultantly}, the estimation of $\boldsymbol{\Phi}(\theta_{r})$ can be achieved by {employing} the least squares
\begin{equation}
\begin{aligned}
\boldsymbol{\breve{\Phi}}(\theta_{r})=\left ( \boldsymbol{J}_{1}\boldsymbol{\widehat{A}}_{r} \right )^{\dagger } \boldsymbol{J}_{2}\boldsymbol{\widehat{A}}_{r}
\end{aligned}   
\end{equation} 
\par {Furthermore}, by performing the singular value decomposition on $\boldsymbol{\breve{\Phi}}(\theta_{r})$, the corresponding eigenvalues $\lambda_{r1},\lambda_{r2},\cdots,\lambda_{rK}$  can be {attained}. So, the  receive elevation angle $\theta_{rk},k=1,2,\cdots,K$  can be estimated as
\begin{equation}
\begin{aligned}
\breve{\theta}_{rk}=arcsin(\frac{\lambda_{rk}}{\pi}),\quad k=1,2,\cdots,K
\end{aligned}   \label{21}
\end{equation}
\par Let $\boldsymbol{\breve{\theta}}=\left [\breve{\theta}_{r1}, \breve{\theta}_{r2},\cdots,\breve{\theta}_{rK} \right ] $.
For the estimated ${\boldsymbol{\breve{Q}}_{r}}$ {through} by using the outer part PARAFAC algorithm, the detailed derivation process for $(\boldsymbol{\tilde{\theta} }_{r},\boldsymbol{\tilde{\phi} }_{r},\boldsymbol{\tilde{\gamma} }_{r},\boldsymbol{\tilde{\eta} }_{r})$ can be {conducted through} the following process.
\par The normalized Poynting vector of ${\boldsymbol{\breve{Q}}_{r}}$  {is}
\begin{equation}
\begin{aligned}
\begin{bmatrix}
\tilde{u}_{r_{k}}\\
\tilde{v}_{r_{k}}\\
\tilde{w}_{r_{k}}\\
\end{bmatrix}\triangleq \frac{\boldsymbol{e}_{r_{k}}}{\left \|\boldsymbol{e}_{r_{k}}  \right \|} \times \frac{\boldsymbol{h}^{*}_{r_{k}}}{\left \|\boldsymbol{h}_{r_{k}}  \right \|}=\begin{bmatrix}
\sin(\tilde{\theta} _{r_{k}})\cos(\tilde{\phi} _{r_{k}})\\
\sin(\tilde{\theta} _{r_{k}})\sin(\tilde{\phi} _{r_{k}})\\
\cos(\tilde{\theta} _{r_{k}})\\
\end{bmatrix}
\end{aligned}  
\end{equation} 
\par Therefore, the estimated transmit elevation angle and azimuth angle $(\tilde{\theta} _{r_{k}},\tilde{\phi} _{r_{k}}),k=1,2,\cdots,K$ can be expressed as
\begin{equation}
\begin{aligned}
\begin{cases}
\tilde{\theta} _{r_{k}} =arctan(\frac{ \tilde{v}_{r_{k}}}{ \tilde{u}_{r_{k}}}) \\
\tilde{\phi} _{r_{k}} =arcsin\left ( \sqrt{\left ( \tilde{v}_{r_{k}} \right )^{2}+ \left ( \tilde{u}_{r_{k}} \right )^{2} } \right )  
\end{cases}\\ 
\end{aligned}   
\end{equation}
\par After obtaining $(\tilde{\theta} _{r_{k}},\tilde{\phi} _{r_{k}})$, the corresponding receive  polarization state vector ${\boldsymbol{g}_{t_{k}}(\gamma_{t_{k}},\eta_{t_{k}})},{ k=1,2,\cdots,K }$ can be expressed as
\begin{equation}
\begin{aligned}
{\boldsymbol{g}_{r_{k}}(\gamma_{r_{k}},\eta_{r_{k}})}&=
\begin{bmatrix}
{\boldsymbol{g}_{1r_{k}}}\\
{\boldsymbol{g}_{2r_{k}}}\\
\end{bmatrix}=[{\boldsymbol{F}(\tilde{\theta} _{r_{k}},\tilde{\phi} _{r_{k}})} ]^{\dagger}\boldsymbol{Q}_{r}	
\end{aligned}   
\end{equation}
\par {Moreover}, the polarization parameters $(\tilde{\gamma} _{r_{k}},\tilde{\eta}_{r_{k}}), k=1,2,\cdots,K$ of the receive EMVS can be expressed as
\begin{equation}
\begin{array}{lcl}
\tilde{\gamma}_{r_{k}} = arctan[\frac{{\boldsymbol{g}_{r_{k}}}}{{\boldsymbol{g}_{r_{k}}}}]  \\
\tilde{\eta}_{r_{k}}=\angle{\boldsymbol{g}_{r_{k}}}	 
\end{array},k=1,2,\cdots,K  
\end{equation}
\par Finally,  the receive 4-D parameters $(\boldsymbol{\theta}_{r},\boldsymbol{\phi}_{r},\boldsymbol{\gamma}_{r},\boldsymbol{\eta}_{r})$  and the transmit 4-D parameters $(\boldsymbol{\theta}_{t},\boldsymbol{\phi}_{t},\boldsymbol{\gamma}_{t},\boldsymbol{\eta}_{t})$ can be {attained by the abovementioned} process.
\subsection{Some Remarks}
${\mathit{Remark\quad 1}}:$ The transmit 4-D parameters $(\boldsymbol{\theta} _{t},\boldsymbol{\phi_{t}},\boldsymbol{\gamma_{t}},\boldsymbol{\eta_{t}})$ are automatically paired, { since} the estimated ${\boldsymbol{\breve{Q}}_{t}}$ and  ${\boldsymbol{\breve{A}}_{t}}$ are {corresponding to each other via the employment of } the inner PARAFAC algorithm.  But, the estimated ${\boldsymbol{\breve{Q}}_{r}}$ by outer PARAFAC algorithm and  ${\boldsymbol{\breve{A}}_{r}}$ by inner PARAFAC algorithm are not one to one correspondance. Thus, the  pairing process for $\boldsymbol{\breve{\theta}}$ estimated by ${\boldsymbol{\breve{A}}_{r}}$ and $(\boldsymbol{\tilde{\theta} }_{r},\boldsymbol{\tilde{\phi} }_{r},\boldsymbol{\tilde{\gamma} }_{r},\boldsymbol{\tilde{\eta} }_{r})$ estimated by ${\boldsymbol{\breve{Q}}_{r}}$ is inevitable. The detailed pairing process can be {carried out} according to   $\textbf{Algorithm 1}$.
\begin{algorithm}  
	\caption{pseudocode for pairing process   }  
	\begin{algorithmic}[1]   
		\Require $\boldsymbol{\breve{\theta}}$ estimated via ${\boldsymbol{\breve{A}}_{r}}$, $\boldsymbol{\tilde{\theta}}$ estimated via ${\boldsymbol{\breve{Q}}_{r}}$
		\Ensure pair matching  $(\boldsymbol{\breve{\theta}_{1}},\boldsymbol{\tilde{\phi} }_{r},\boldsymbol{\tilde{\gamma} }_{r},\boldsymbol{\tilde{\eta} }_{r})$
		\State $\boldsymbol{Index} = zeros(1,K) $ 
		\State $\boldsymbol{for} \quad i=1:K  $ 
		\State  \quad $\boldsymbol{for} \quad j=1:K  $   
		\State  \quad \quad \quad $\boldsymbol{Index}(i)=\underset{j}{{argmin}} \left | \boldsymbol{\tilde{\theta}}(i)-\boldsymbol{\breve{\theta}}(j) \right |$
		\State $\boldsymbol{\breve{\theta}_{1}}=\boldsymbol{\breve{\theta}}[\boldsymbol{Index}]$
		\State \Return{$(\boldsymbol{\breve{\theta}_{1}},\boldsymbol{\tilde{\phi} }_{r},\boldsymbol{\tilde{\gamma} }_{r},\boldsymbol{\tilde{\eta} }_{r})$}  
	\end{algorithmic}  
\end{algorithm} 
\par {Afterwards},  the pair matching receive 4-D parameters $(\boldsymbol{\breve{\theta}_{1}},\boldsymbol{\tilde{\phi} }_{r},\boldsymbol{\tilde{\gamma} }_{r},\boldsymbol{\tilde{\eta} }_{r})$ can be {achieved, in which case,} the transmit 4-D parameters and the receive 4-D parameters are also paired. 
\par ${\mathit{Remark\quad 2}}$: Based on the Kruskal's condition in \cite{GKolda}-\cite{WRao}, the uniqueness of the outer and inner PARAFAC decomposition can be guaranted by 
\begin{equation}
\begin{aligned}
\left\{\begin{matrix}
2K+2 \le \kappa_{\boldsymbol{A}_{tqr}}+\kappa_{\boldsymbol{Q}_{r}} + \kappa_{\boldsymbol{S}}  \\
2K+2 \le\kappa_{\boldsymbol{A}_{t}}+\kappa_{\boldsymbol{Q}_{t}} + \kappa_{\boldsymbol{A}_{r}}
\end{matrix}\right.
\end{aligned}   
\end{equation}
\par And, accodring to (\ref{18}), $K$ also {relies} on the rotation invariant relationship when {estimating} the transmit/receive elevation angle, namely
\begin{equation}
\begin{aligned}
\left\{\begin{matrix}
K\leq M-1  \\
K\leq N-1
\end{matrix}\right.
\end{aligned}   
\end{equation}
Thus, the maximum resolvable targets $K = min\left [ M-1,N-1 \right ] $.
\par ${\mathit{Remark\quad 3}}$: The computational complexity of the proposed nested-PARAFAC algorithm {is mainly determined by the} outer alternating ALS and inter alternating ALS. Thus, the  total computational complexity is $o \left ( \kappa\left [ \left ( 6MN+M+N+L+12 \right )K^{2}  \right ]  \right ) $.
\section{Simulation Results}
\par
\begin{figure}[htbp]
	\centering
	\subfigure[angle parameters]{
		\includegraphics[width=7.0cm]{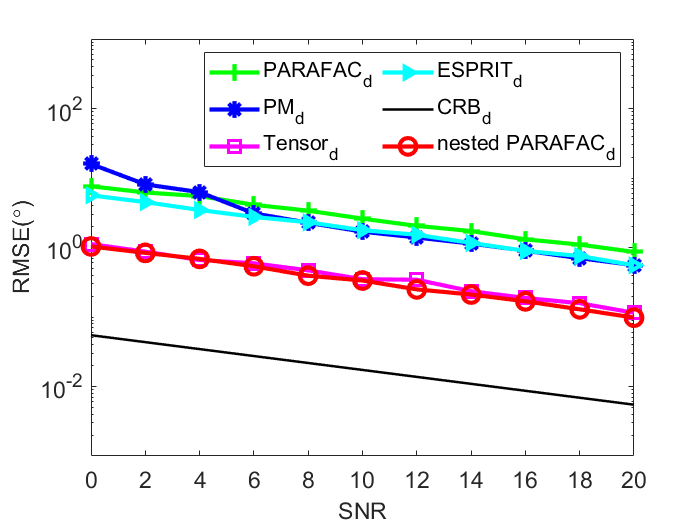}
	}
	\quad
	\subfigure[ polarization paramters]{
		\includegraphics[width=7.0cm]{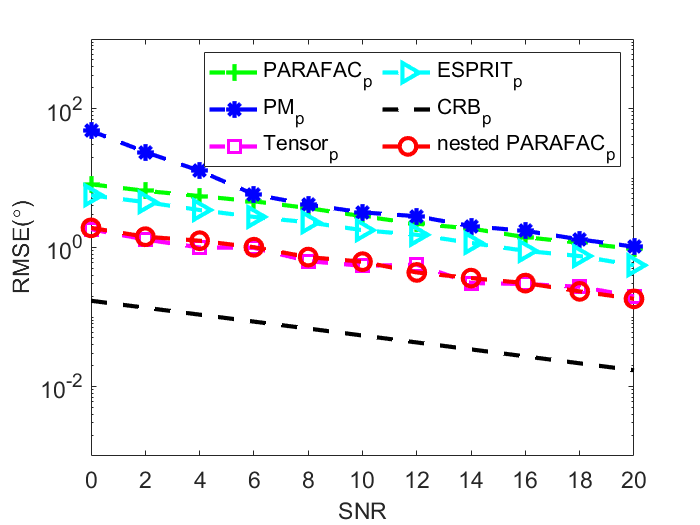}
	}
	\caption{RMSE performance of different algorithms versus SNR }\label{figsnr}
\end{figure}

 Here, the RMSEs performance of different algorithms versus SNR are compared. The ESPRIT algorithm in \cite{Chintagunta}, the PM algorithm in \cite{LiuT}, the Tensor subspace-based algorithm in \cite{MaoC} and the PARAFAC algorithm in \cite{WenFShi} are {considered for comparison.}. The average root mean square error { is calculated as}  $RMSE = \sqrt{\frac{1}{100K}\sum_{i=1}^{100}\|\hat{\boldsymbol{\alpha}}-\boldsymbol{\alpha}\|^{2} }$, where $\boldsymbol{\alpha}$ and $\hat{\boldsymbol{\alpha}}$ denote the {true} angle parameters and estimated angle parameters, respectively. The number of transimit MEVS and receive EMVS are set to $9$ and $10$, respectively. Assume that there are $K=3$ far-field targets with $\boldsymbol{\theta}_{t}=[40^{\circ},20^{\circ},30^{\circ}]$,  $\boldsymbol{\phi}_{t}=[15^{\circ},25^{\circ},35^{\circ}]$, $\boldsymbol{\gamma}_{t}=[10^{\circ},22^{\circ},35^{\circ}]$, $\boldsymbol{\eta}_{t}=[38^{\circ},48^{\circ},56^{\circ}]$, $\boldsymbol{\theta}_{r}=[24^{\circ},38^{\circ},16^{\circ}]$, $\boldsymbol{\phi}_{r}=[21^{\circ},32^{\circ},55^{\circ}]$, $\boldsymbol{\gamma}_{r}=[42^{\circ},33^{\circ},60^{\circ}]$, $\boldsymbol{\eta}_{r}=[17^{\circ},27^{\circ},39^{\circ}]$, respectively. The SNR increases from $0dB$ to $20dB$ and the snapshot is set to $200$ for different {SNRs}. The suffix '-d'and '-p' in the legend refer to angle parameter and polarization parameter, respectively. Fig. \ref{figsnr} {shows} that the proposed algorithm exhibits {an excellent} 8D parameters estimation performance compared with the state-of-art algorithms {on} the conditions of different {SNRs}.
\section{Conclusion}
In this letter, {a} high accuracy estimation for the transmit 4D parameters and the receive 4D parameters in the bistatic EMVS-MIMO { radar has been investigated.} By utilizing the tensor structure,  the array receive data are rearranged into an outer part three way tensor models and an inner part {three} way tensor models, which are resolved by the proposed nested PARAFAC algorithm. The proposed nested PARAFAC algorithm can effectively solve the transmit/receive elevation angle, transmit/receive azimuth angle, transmit/ receive polarization angle and transmit/ receive polarization phase difference. More significant is that the proposed algorithm provides a new strategy to deal with the parameters estimaion problem for other signal models similar to the one in this letter. In {the upcoming} studies, we will consider how to apply the proposed nested PARAFAC algorithm to the bistatic EMVS-MIMO radar with sparse transmit EMVS and sparse receive EMVS.

\section*{References}
\def\refname{}

\end{document}